# Imaging of non tumorous and tumorous human brain tissue with full-field optical coherence tomography

Osnath Assayag*[1], Kate Grieve*[1], Bertrand Devaux[2,3], Fabrice Harms[1], Johan Pallud[2,3], Fabrice Chretien[2,3], Claude Boccara[1] and Pascale Varlet[2,3]

*equal first authors

[1] Inserm U979 "Wave Physics For Medicine" ESPCI –ParisTech – Institut Langevin, 1 rue Jussieu, 75005 France, [2] Centre Hospitalier Sainte-Anne, 1 rue Cabanis 75014 Paris, France, [3] University Paris Descartes, France.

**Keywords:** optical imaging, digital pathology, brain imaging, brain tumor, glioma

**Corresponding Author:** Kate Grieve, ESPCI – Institut Langevin, 1 rue Jussieu, 75005 France, kate.grieve@espci.fr, +33 (0)1 82 72 61 28

**Abbreviations:** FF-OCT: full field optical coherence tomography, OCT: optical coherence tomography

**Abstract**

A prospective study was performed on neurosurgical samples from 18 patients to evaluate the use of Full-Field Optical Coherence Tomography (FF-OCT) in brain tumor diagnosis.

FF-OCT captures en face slices of tissue samples at 1µm resolution in 3D with a typical 200µm imaging depth. A 1cm² specimen is scanned at a single depth and processed in about 5 minutes. This rapid imaging process is non-invasive and requires neither contrast agent injection nor tissue preparation, which makes it particularly well suited to medical imaging applications.

Temporal chronic epileptic parenchyma and brain tumors such as meningiomas, low-grade and high-grade gliomas, and choroid plexus papilloma were imaged. A subpopulation of neurons, myelin fibers and CNS vasculature were clearly identified. Cortex could be discriminated from white matter, but individual glial cells as astrocytes (normal or reactive) or oligodendrocytes were not observable.

This study reports for the first time on the feasibility of using FF-OCT in a real-time manner as a label-free non-invasive imaging technique in an intra-operative neurosurgical clinical setting to assess tumorous glial and epileptic margins.

# 1   Introduction

## *1.1   Primary CNS tumors*

Primary central nervous system (CNS) tumors represent a heterogeneous group of tumors with benign, malignant and slow-growing evolution. In France, annual incidence of primary CNS tumors is 5000 new cases (Rigau et al 2011). Despite considerable progress in diagnosis and treatment, the survival rate following a malignant brain tumor remains low and there are a reported 3000 annual deaths from CNS tumors in France (INCa 2011). Overall survival from brain tumors depends on the complete resection of the tumor mass, as defined on postoperative imaging, and associated with updated adjuvant radiation therapy and chemotherapy regimen for malignant tumors (Soffietti et al 2010). Therefore, there is a need for evaluating the completeness of the tumor resection at the end of the surgical procedure, as well as an intraoperative identification of the different components of the tumor, i.e. tumor tissue, necrosis, infiltrated parenchyma (Kelly et al 1987). In particular, the persistence of non visible tumor tissue or isolated tumor cells infiltrating brain parenchyma may lead to additional resection.

For low-grade tumors located close to eloquent brain areas, a maximal safe resection sparing functional tissue warrants the current use of intraoperative techniques that guide a more complete tumor resection. During awake surgery, speech or fine motor skills are monitored, while cortical and subcortical stimulations are performed to identify functional areas (Sanai et al 2008). Intraoperative MRI provides images of the surgical site as well as tomographic images of the whole brain that are sufficient for an approximate evaluation of the abnormal excised tissue, but offer low resolution (typically 1 to 1.5 mm) and produce artifacts at the air-tissue boundary of the surgical site.

Histological and immunohistochemical analysis of neurosurgical samples remains the current gold standard method used to analyze tumorous tissue. However, these methods require long (2 to 3 day) multiple steps, and use of carcinogenic chemical products that would not be technically possible intra-operatively. To obtain extemporaneous information, intra-operative cytological smear tests are performed. However the information on the tissue architecture is thereby lost and the analysis is carried out on only a limited area of the sample (1mm x 1mm).

Intraoperative optical imaging techniques are recently developed high resolution imaging modalities that may help the surgeon to identify the persistence of tumor tissue at the resection boundaries. Using a conventional operating microscope, illuminating the operative field by a Xenon lamp, gives an overall view of the surgical site but is limited by the poor discriminative capacity of the white light illumination at the surgical site interface. Better discrimination between normal and tumorous tissue has been obtained using fluorescence properties of tumor cells labeled with 5-ALA administered preoperatively. Tumor tissue shows a strong ALA-induced PPIX fluorescence at 635 and 704 nm when the operative field is illuminated with a 440nm-filtered lamp. More complete resections of high grade gliomas have been demonstrated using 5-ALA fluorescence guidance (Stummer et al 2000), however brain parenchyma infiltrated by isolated tumor cells is not fluorescent, reducing the interest of this technique when resecting low-grade gliomas.

Refinement of this induced fluorescence technique has been obtained using a confocal microscope and intraoperative injection of sodium fluorescein. A 488 nm laser light illuminates the operative field and a tissue contact analysis is performed using a handheld surgical probe (field of view less than 0.5 x 0.5 mm) which scans the fluorescence of the surgical interface at the 505 – 585 nm band. Fluorescent isolated tumor cells are clearly identified at a depth from 0 to 500µm from the resection border (Sanai et al 2011), potentially making this technique interesting in low-grade glioma resection.

100 On review of the state-of-the-art, a need is identified for a quick and reliable method of providing the neurosurgeon with architectural and cellular information without the need for injection or oral intake of exogenous markers in order to guide the neurosurgeon and optimize surgical resections.

*1.2    Full-Field Optical Coherence Tomography:*

Introduced in the early 1990s (Huang et al. 1991), Optical Coherence Tomography (OCT) uses interference to precisely locate light deep inside tissue. The photons coming from the small volume of interest are distinguished from light scattered by the other parts of the sample by the use of an interferometer and a light source with short
110 coherence length. Only the portion of light with the same path length as the reference arm of the interferometer, to within the coherence length of the source (typically a few µm), will produce interference. A two-dimensional B-scan image is captured by scanning. Recently, the technique has been improved, mainly in terms of speed and sensitivity, through spectral encoding.

A new OCT technique called Full-Field Optical Coherence Tomography (FF-OCT) enables both a large field of view and high resolution over the full field of observation (Dubois et al 2002, 2004). This allows navigation across the wide field image to follow the morphology at different scales and different positions. FF-OCT uses a simple halogen or light-emitting diode (LED) light source for full field illumination, rather than expensive, bulky lasers and point-by-point scanning components required for conventional OCT. The illumination level is low enough to maintain the sample integrity: the power incident on the sample is less than 1mW/mm$^2$ using deep red and near infrared light. FF-OCT provides the highest OCT 3D resolution of 1.5 x 1.5 x 1 µm$^3$ (X x Y x Z) on unprepared label-free tissue samples down to approximately 200µm - 300µm depth over a large field of view that allows digital zooming down to the cellular level. Interestingly, it produces en face images in the native field view (rather than the cross-sectional images of conventional OCT), which mimic the histology process, thereby facilitating the reading of images by pathologists. Moreover, as for conventional OCT, it does not require tissue slicing or modification of any kind (no tissue fixation, coloration, freezing, paraffin embedding). FF-OCT image acquisition and processing time is less than five minutes for a typical 1cm$^2$ sample (Assayag et al 2013) and the imaging performance has been shown to be equivalent in fresh or fixed tissue (Assayag et al 2013, Dalimier and Salomon 2012). In addition, FF-OCT intrinsically provides digital images suitable for telemedicine.

Numerous studies have been published over the past two decades demonstrating the suitability of OCT for in vivo or ex vivo diagnosis. OCT imaging has been previously applied in a variety of tissues such as the eye (Swanson et al 1993, Grieve et al 2004), upper aerodigestive tract (Betz et al 2008, Ozawa et al 2009,

Chen et al 2007), gastrointestinal tract (Tearney et al 1998), and breast tissue and lymph nodes (Adie and Boppart 2009, Boppart et al 2004, Hsiung et al 2007, Luo et al 2005, Nguyen et al 2009, Zhou et al 2010, Zysk and Boppart 2006).

In the CNS, published studies that evaluate OCT (Bizheva et al 2005, Böhringer et al 2006, 2009, Boppart et al 1998, Boppart 2003) using time-domain (TD) or spectral domain (SD) OCT systems had insufficient resolution (10 to 15µm axial) for visualization of fine morphological details. A study of 9 patients with gliomas carried out using a TD-OCT system led to classification of the samples as malignant versus benign (Böhringer et al 2009). However, the differentiation of tissues was achieved by considering the relative attenuation of the signal returning from the tumorous zones in relation to that returning from healthy zones. The classification was not possible by real recognition of CNS microscopic structures. Another study showed images of brain microstructures obtained with an OCT system equipped with an ultra-fast laser that offered axial and lateral resolution of 1.3µm and 3µm respectively (Bizheva et al 2005). In this way, it was possible to differentiate malignant from healthy tissue by the presence of blood vessels, microcalcifications and cysts in the tumorous tissue. However the images obtained were small (2mm x 1mm), captured on fixed tissue only and required use of an expensive large laser thereby limiting the possibility for clinical implementation.

This study is the first to analyze non-tumorous and tumorous human brain tissue samples using FF-OCT.

## 2      Materials and Methods

*2.1    Instrument*

The experimental arrangement of FF-OCT (Figure 1A) is generally based on a configuration that is referred to as a Linnik interferometer (Dubois et al 2002). A halogen lamp is used as a spatially incoherent source to illuminate the whole field of an immersion microscope objective. The signal is extracted from the background of incoherent backscattered light using a phase-shifting method implemented in custom-designed software. This study was performed on a commercial FF-OCT system (LightCT, LLTech, France).

Capturing "en face" images allows easy comparison with histological sections. The resolution, pixel number and sampling requirements result in a native field of view that is limited to about 1mm$^2$.  The sample is moved on a high precision mechanical platform and a number of fields are stitched together to display a significant field of view (Beck et al 2000). The FF-OCT microscope is housed in a compact setup (Figure 1B) that is about the size of a standard optical microscope (310x310x800mm L x W x H).

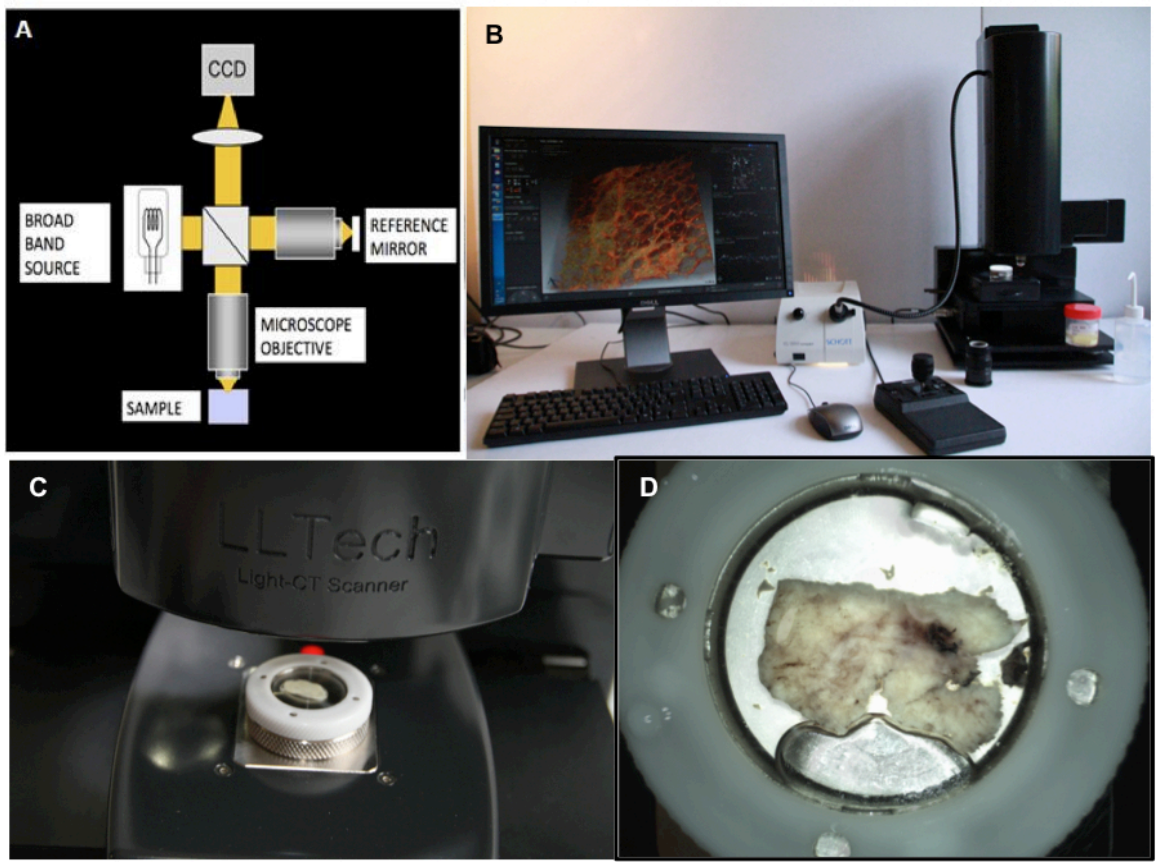

Figure 1: system schematic (A), photograph (B), sample holder close up (C), sample close up (D)

## 2.2 Imaging Protocol

All images presented in this study were captured on fresh brain tissue samples from patients operated on at the Neurosurgery Department of Sainte-Anne Hospital, Paris. Informed and written consent was obtained in all cases following the standard procedure at Sainte-Anne Hospital from patients who were undergoing surgical intervention. Fresh samples were collected from the operating theater immediately after resection and sent to the pathology department. A pathologist dissected each sample to obtain a 1-2cm$^2$ piece and made a macroscopic observation to orient the specimen in order to decide on which side to image. The sample was immersed in physiological serum, placed in a cassette, numbered, and brought to the FF-OCT imaging facility in a nearby laboratory (15 minutes distant) where the FF-OCT images were captured. The sample was placed in a custom holder with a coverslip on top (Figures 1C, 1D). The sample was raised on a piston to rest gently against the coverslip in order to flatten the surface and so optimize the image capture. The sample is automatically scanned under a 10X 0.3 numerical aperture (NA) immersion microscope objective. The entire area of each sample was imaged at a depth of 20μm beneath the sample surface. This depth has been reported to be optimal for comparison of FF-OCT images to histology images in a previous study on breast tissue (Assayag et al 2013). Once FF-OCT imaging was done, each sample was immediately fixed in formaldehyde and returned to the pathology department where it underwent standard processing in order to compare the FF-OCT images to histology slides.

## 3    Results

18 samples from 18 adult patients (4 male, 14 female) of age range 19-81 years have been included in the study: 1 mesial temporal lobe epilepsy and 1 cerebellum adjacent to a pulmonary adenocarcinoma metastasis (serving as the non-tumor brain samples), 7 diffuse supratentorial gliomas (4 WHO gradeII, 3 WHO grade III), 5 meningiomas, 1 hemangiopericytoma, and 1 choroid plexus papilloma. Patient characteristics are detailed in Table 1.

*3.1    FF-OCT imaging identifies myelinated axon fibers, neuronal cell bodies and vasculature in the human epileptic brain and cerebellum*

The cortex and the white matter are clearly distinguished from one another (Figure 2). Indeed, a subpopulation of neuronal cell bodies (Figures 2B, 2C) as well as myelinated axon bundles leading to the white matter could be recognized (Figure 2D, 2E). Neuronal cell bodies appear as black triangles (Figure 2C). Myelinated axons are numerous, well discernible as small fascicles and appear as bright white lines (Figure 2E). As the cortex does not contain many myelinated axons, it appears dark grey. Brain vasculature is visible (Figures 2F and 2G), and small vessels are distinguished by a thin collagen membrane that appears light grey. Of note, based on myelinated axon diameter, we show that FF-OCT lateral resolution is better than 1.5µm, this value being in accordance with its theoretical value of about 1µm. Video 1 in supplementary material shows a movie composed of a series of en face 1µm thick optical slices captured over 100µm into the depth of the cortex tissue. The myelin fibers and neuronal cell bodies are seen in successive layers.

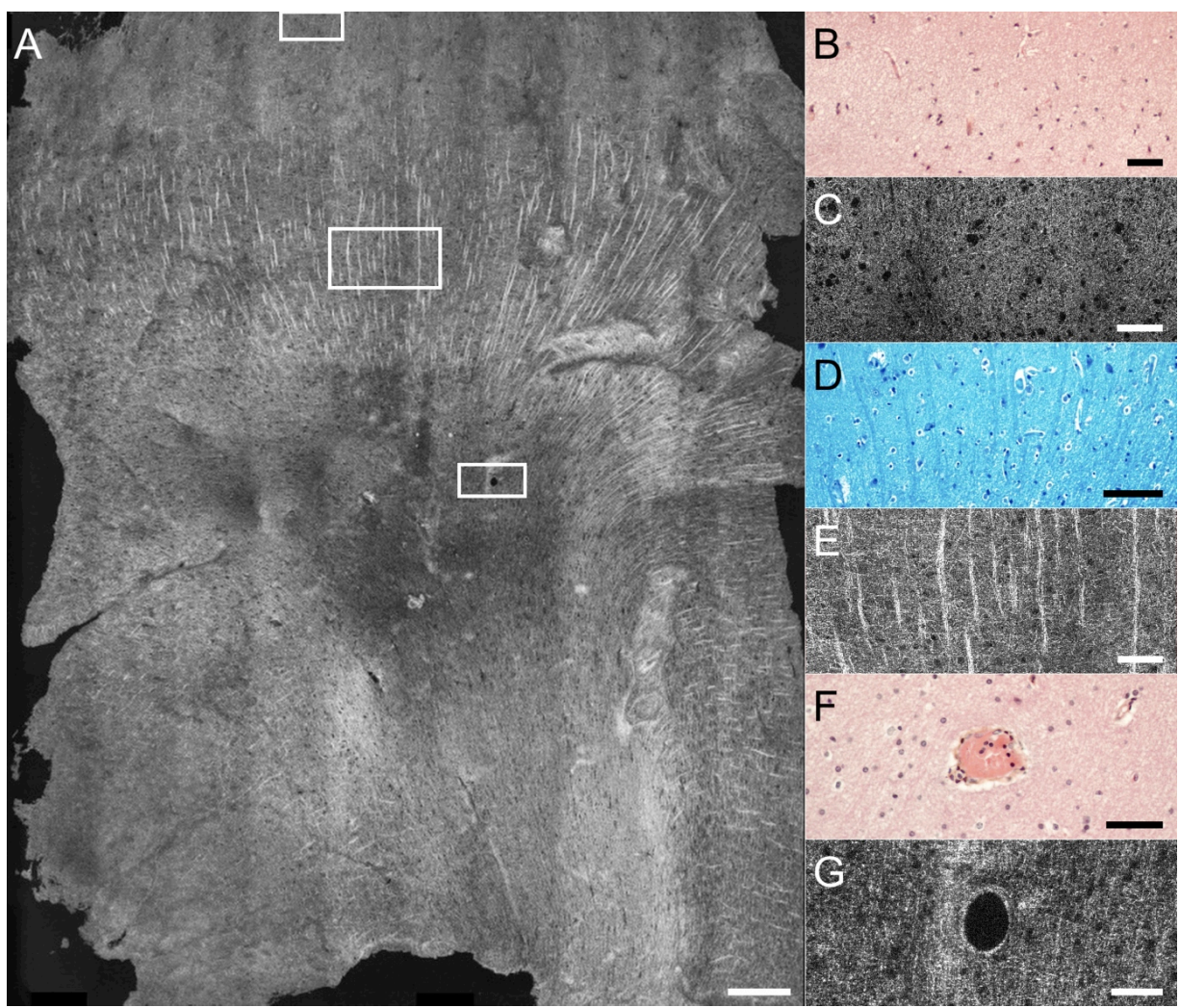

Figure 2: Cortex is distinguished from white matter. Cortex appears grey. B-C neuronal cell bodies, D-E myelinated axon bundles leading to white matter, F-G vasculature. B and F Hemalun and phloxin stainings and D Luxol blue staining. Rectangles indicate locations of zooms: top = C, middle = E, bottom = G. Scale bars show 500µm (A), 50µm (B, C, F, G) and 80µm (D, E).

240

The different regions of the human hippocampal formation are easily recognizable (Figure 3). Indeed, CA1 field and its stratum radiatum, CA4 field, the hippocampal fissure, the dentate gyrus, and the alveus are easily distinguishable. Other structures become visible by digital zooming in on the FF-OCT image. The large pyramidal neurons of the CA4 field (Figure 3B) and the granule cells that constitute the stratum granulosum of the dentate gyrus are visible, as black triangles and as small round dots, respectively (Figure 3D).

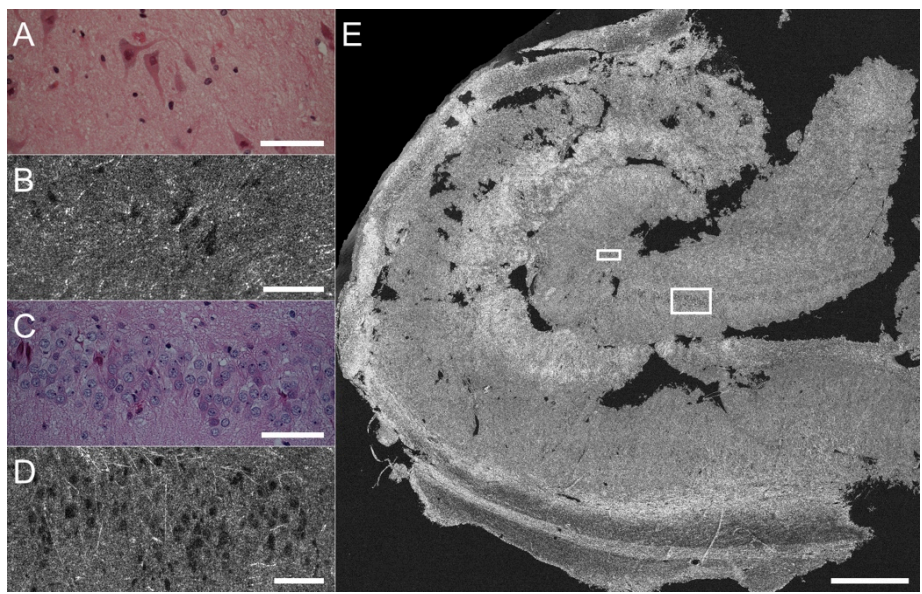

Figure 3: Hippocampus. CA1 field and stratum radiatum, CA4 field, the hippocampal
250 fissure, the dentate gyrus, and the alveus are distinguished. A-B pyramidal neurons of CA4, C-D granular cells constitute the stratum granulosum of the dentate gyrus. A and C Hemalun and phloxin stainings. Rectangles indicate locations of zooms: top = B, bottom = D. Scale bars show 40μm (A, B), 80μm (C, D), 900μm (E).

In the normal cerebellum, the lamellar or foliar pattern of alternating cortex and central white matter is easily observed (Figure 4A). By numerical zooming, Purkinje and granular neurons also appear as black triangles or dots, respectively (Figure 4C), and myelinated axons are visible as bright white lines (Figure 4E).

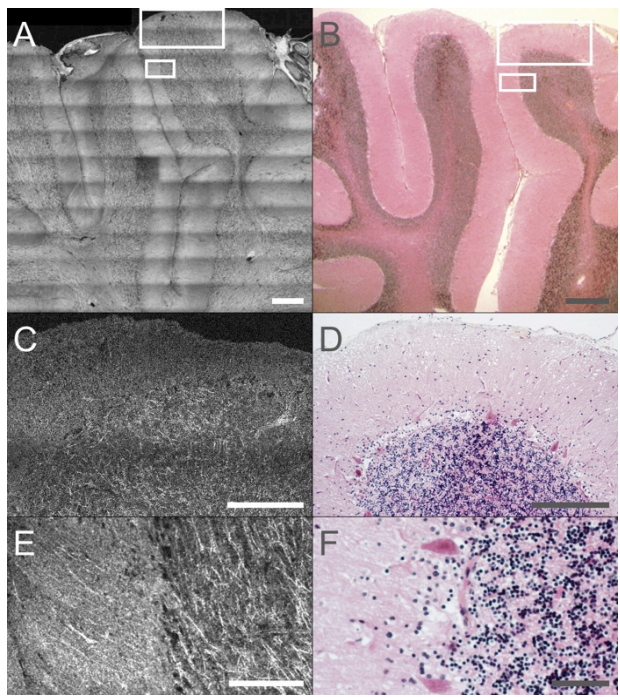

260

Figure 4: Cerebellum. A-B: the lamellar or foliar pattern of alternating cortex and central white matter. C-D, E-F zooms show cerebellar cortex and granular layer, in E-F Purkinje and granular neurons are distinguished as black triangles or dots, respectively, and myelinated axons as bright white lines. B, D and F Hemalun and phloxin stainings. Rectangles indicate locations of zooms: top = C-D, bottom = E-F. Scale bars show 500μm (A, B), 350μm (C, D), 100μm (E, F)

*3.2 FF-OCT images distinguishes meningiomas from haemangiopericytoma in meningeal tumors*

The classical morphological characteristics of a meningioma are visible on the FF-OCT image: large lobules of tumorous cells appearing in light grey (Figure 5A), demarcated by collagen-rich bundles (Figure 5B), which are highly scattering and appear a brilliant white in the FF-OCT images. The classical concentric tumorous cell clusters (whorls) are very clearly distinguished on the FF-OCT image (Figure 5D). In addition the presence of numerous cell whorls centered with central calcifications (psammoma bodies) is revealed (Figure 5F). Collagen balls appear bright white on the FF-OCT image (Figure 5H) and calcifications appear black generating a target-like image (Figure 5H).

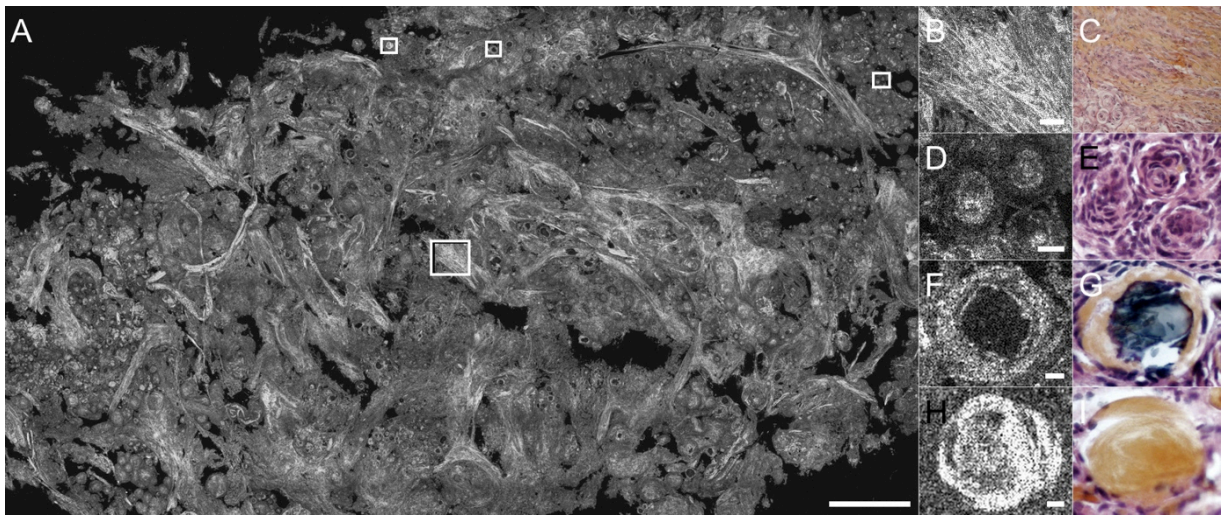

Figure 5: Meningioma psammoma. B-C collagen bundles, D-E whorls, F-G calcifications, H-I collagen balls. C, E, G and I Hemalun and phloxin stainings. Rectangles indicate locations of zooms: top left = H, top middle = F, top right = D, bottom = B. Scale bars show 500µm (A), 50µm (B,C), 10µm (D, E, F, G, H, I).

Mesenchymal non-meningothelial tumors such as hemangiopericytomas represent a classic differential diagnosis of meningiomas. In FF-OCT, the hemangiopericytoma shows a more monotonous appearance than meningiomas, with a highly vascular branching component with starghorn-type vessels (Figure 6A, 6C).

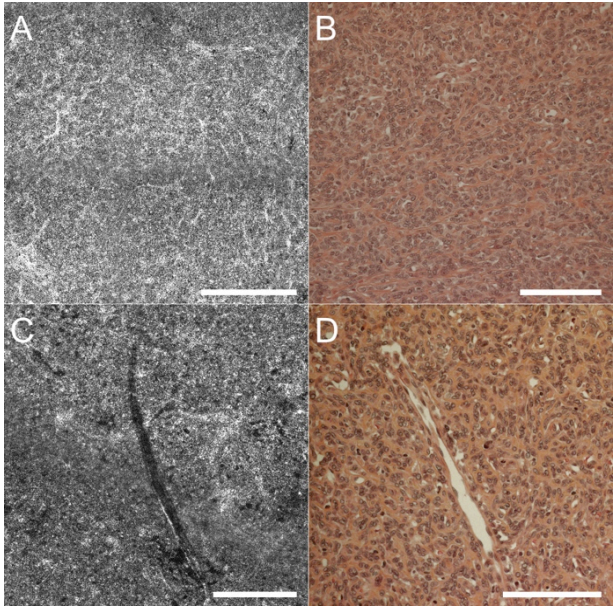

Figure 6: Hemangiopericytoma. A-B collagen network and branching vascular space. Starghorn sinusoids appear white. C-D vessel. B and D Hemalun and phloxin stainings. Scale bars show 250µm.

### 3.3  FF-OCT images identifies choroid plexus papilloma

The choroid plexus papilloma appears as an irregular coalescence of multiple papillas composed of elongated fibrovascular axes covered by a single layer of choroid glial cells (Figure 7). By zooming in on an edematous papilla, the axis appears as a black structure covered by a regular light grey line (Figure 7B). If the papilla central axis is hemorrhagic, the fine regular single layer is not distinguishable (Figure 7C). Additional digital zooming in on the image reveals cellular level information and some nuclei of plexus choroid cells can be recognized. However, cellular atypia or mitosis are not visible. These represent key diagnosis criteria used to differentiate choroid plexus papilloma (grade I) from atypical plexus papilloma (grade II).

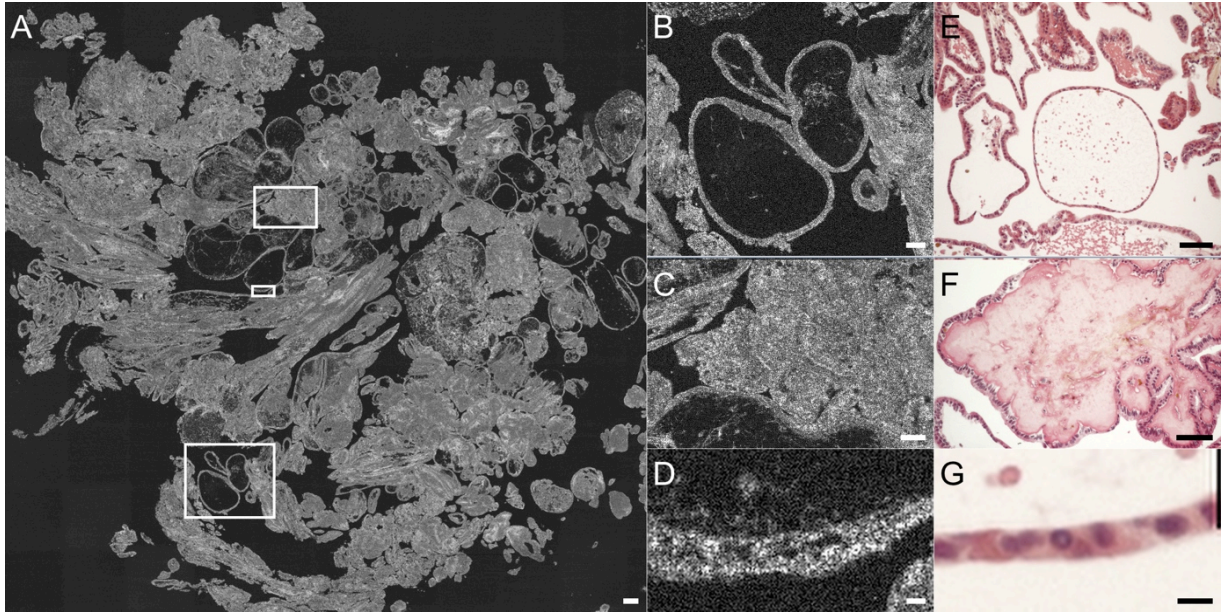

Figure 7: Papilloma – cauliflower-like aspect. B-E empty papilla, C-D blood filled papilla, D-G single layer of plexus cells. E, F and G Hemalun and phloxin stainings. Rectangles indicate locations of zooms: top = C, middle = D, bottom = B. Scale bars show 150µm (A), 50µm (B, C, E, F), 20µm (D, G).

### 3.4 *FF-OCT images detect the brain tissue architecture modifications generated by diffusely infiltrative gliomas*

Contrary to the choroid plexus papillomas which have a very distinctive architecture in histology (cauliflower-like aspect), very easily recognized in the FF-OCT images (Figure 7A to G), diffusely infiltrating glioma do not present a specific tumor architecture (Figure 8) as they diffusely permeate the normal brain architecture. Hence, the tumor glial cells are largely dispersed through a quite nearly normal brain parenchyma (Figure 8E). In our experience, diffuse low-grade gliomas (less than 20% of tumor cell density) are not recognized on FF-OCT images. The presence of infiltrating tumorous glial cells attested by high magnification histological observation (irregular atypical cell nuclei compared to normal oligodendrocytes) is not detectable with the current generation of FF-OCT devices. However, in high-grade gliomas, (Figure 8G-K), the loss of normal parenchyma architecture is easily observed.

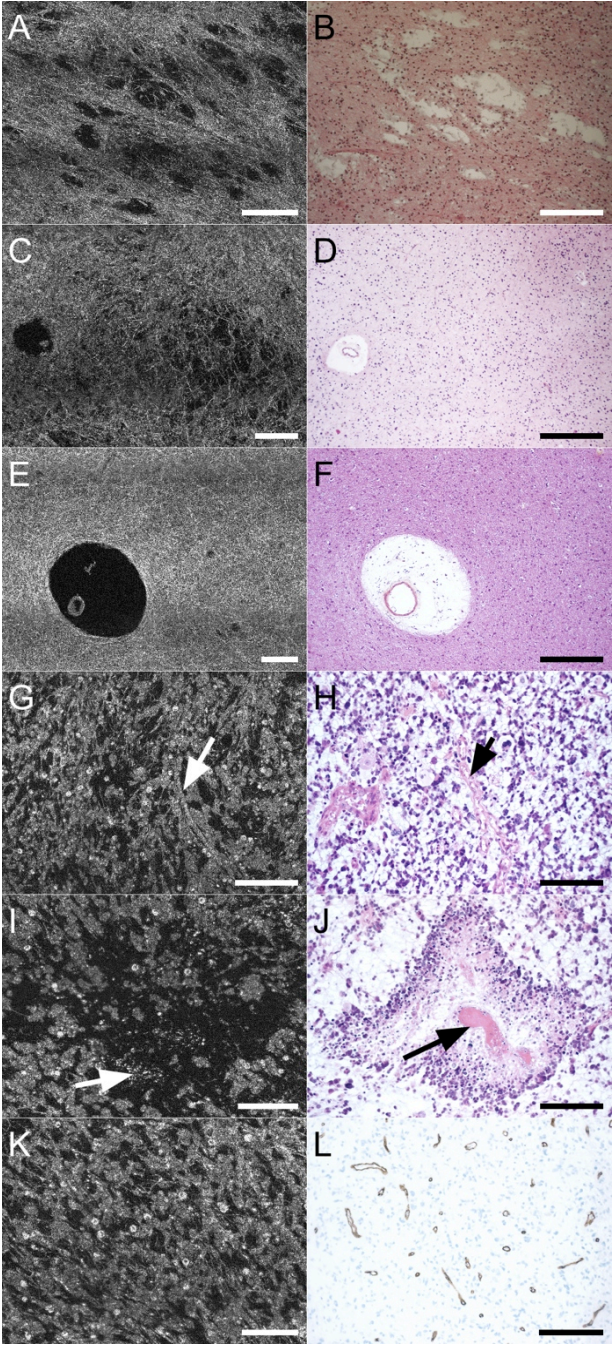

Figure 8: Glioma. Three different cases are shown here (A-B; C-F; G-L). A-B microcysts in an oligo-astrocytoma grade 2; C-D microcystic areas in an astrocytoma

grade 2; E-F: Enlarged Virchow-Robin spaces in an astrocytoma grade 2; G-H microvessels (arrow) and tumorous glial cells in an oligo-astrocytoma grade 3; I-J pseudo-palisading necrosis in an oligo-astrocytoma grade 3. Necrosis appears as dark diamond shaped area. White powdery substance in center of dark space (white arrow) is lysed cells (necrotic cells/centers). Dark arrow on histology shows a vessel. K-L vasculature in an oligo-astrocytoma grade 3 is immediately visible in white in FF-OCT images, while in histology an additional coloration is required to visualize this feature. B, D, F, H and J Hemalun and phloxin stainings and CD34 immunostaining (L). Scale bars show 250µm (A, B), 100µm (C-F), 20µm (G, H), 10µm (I-L).

## 4   Discussion

We present here the first large size images (i.e. on the order of 1-3 cm²) acquired using an OCT system that offer a spatial resolution comparable to histological analysis, sufficient to distinguish microstructures of the human brain parenchyma. Firstly, the FF-OCT technique and the images presented here combine several practical advantages. The imaging system is compact, it can be placed in the operating room, the tissue sample does not require preparation and image acquisition is rapid. This technique thus appears as a promising tool suitable intraoperatively to help neurosurgeons and pathologists.

Secondly, resolution is sufficient (on the order of 1µm axial and lateral) to distinguish brain tissue microstructures. Indeed, it was possible to distinguish neuron cell bodies in the cortex and axon bundles going toward white matter. Myelin fibers of 1µm in diameter are visible on the FF-OCT images. Thus FF-OCT may serve as a real-time anatomical locator.

In addition, histological architectural characteristics of meningothelial, fibrous, transitional and psammomatous meningiomas were easily recognizable on the FF-OCT images (lobules and whorl formation, collagenous-septae, calcified psammoma bodies, thick vessels). In this way, psammomatous and transitional meningiomas presented distinct architectural characteristics in FF-OCT images in comparison to those observed in hemangiopericytoma. Thus, FF-OCT may serve as an intraoperative tool, in addition to extemporaneous examination, to refine differential diagnosis between pathological entities with different prognoses and surgical managements.

Diffuse glioma was essentially recognized by the loss of normal parenchyma architecture. However, FF-OCT images allowed glioma detection only if the glial cell density is greater than 20%. The FF-OCT technique is therefore not currently suitable for the evaluation of low tumorous infiltration or tumorous margins. This evaluation at the individual tumor cell level is only possible by IDH1R132 immunostaining in IDH1 mutated gliomas in adults (Preusser et al 2011). One of the current limitations of the FF-OCT technique for use in diagnosis is the difficulty in estimating the nuclear/cytoplasmic boundaries, the size and form of nuclei as well as the nuclear-cytoplasmic ratio of cells. This prevents precise classification into tumor subtypes and grades.

To increase the accuracy of diagnosis of tumors where cell density measurement is necessary for grading, perspectives for the technique include development of a multimodal system (Harms et al 2012) to allow simultaneous co-localized acquisition of FF-OCT and fluorescence images. Use of this multimodal setup may be of interest in further exploration of the possibility of diagnosis and tumor grading direct from optical images, as the fluorescence channel images would show cell nuclei.

380 However, the use of contrast agents for the fluorescence channel means that the multimodal imaging technique is no longer non invasive, and this may be undesirable if the tissue is to progress to histology following optical imaging.

In its current form therefore, FF-OCT is not intended to serve as a diagnostic tool, but should rather be considered as an additional intraoperative aid in order to determine in a short time whether or not there is suspicious tissue present in a sample. It does not aim to replace histological analyses but rather to complement them, by offering a tool at the intermediary stage of intraoperative tissue selection. In a few minutes, an image is produced that allows the surgeon or the pathologist to assess the content of
390 the tissue sample. The selected tissue, once imaged with FF-OCT, may then proceed for conventional histology procedures in order to obtain the full diagnosis (Assayag et al 2013, Dalimier and Salomon 2012).

Development of FF-OCT to allow in vivo imaging is underway, and first steps include increasing camera acquisition speed. First results of in vivo rat brain imaging have been achieved with an FF-OCT prototype setup, and show real-time visualization of myelin fibers (Ben Arous et al 2011) and movement of red blood cells in vessels (Binding et al 2011). To respond more precisely to surgical needs, it would be preferable to integrate the FF-OCT system into a surgical probe. Work in this
400 direction is currently underway and preliminary images of skin and breast tissue have been captured with a rigid probe FF-OCT prototype (Latrive and Boccara 2011).

In conclusion, we have demonstrated the capacity of FF-OCT for imaging of human brain samples. This technique has potential as an intraoperative tool for determining tissue architecture and content in a few minutes. The 1µm$^3$ resolution and wide field down to cell level views offered by the technique allowed identification of features of non-tumorous and tumorous tissue such as myelin fibers, neurons, microcalcifications, tumor cells, microcysts, and blood vessels. Correspondence with histological slides was good, allowing use in clinical practice for tissue selection for

410    biobanking for example. Future work to extend the technique to in vivo imaging by rigid probe endoscopy is underway.

**Acknowledgements:** The authors wish to thank LLTech SAS for use of the LightCT Scanner.

**Supplementary material**

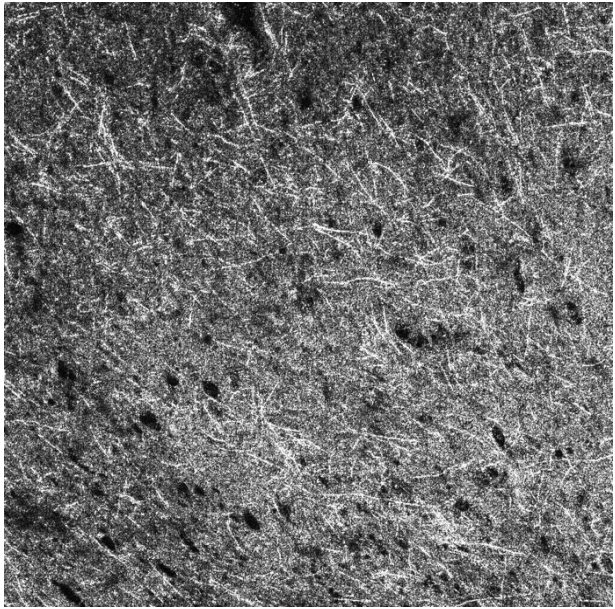

Video 1 in shows a movie composed of a series of en face 1µm thick optical slices captured over 100µm into the depth of the cortex tissue. The myelin fibers and neuronal cell bodies are seen in successive layers. Field size is 800µm x 800µm.